\documentclass{aa}

\usepackage{hyperref}
\makeatletter
\def\HyPsd@expand@utfvii{}
\makeatother

\usepackage[varg]{txfonts}
\usepackage{xcolor}
\usepackage[normalem]{ulem}
\usepackage{comment}
\usepackage{eht}

\makeatletter
\@ifpackageloaded{linenoaa}{\let\linenumbers\relax}{}
\makeatother

\DeclareRobustCommand{\okina}{%
  \raisebox{\dimexpr\fontcharht\font`A-\height}{%
    \scalebox{0.8}{`}%
  }%
}

\begin{document}
\makeatletter

\title{Black hole accretion and radiation variability in GRMHD simulations with Rezzolla-Zhidenko spacetime}

\author{
  Kotaro Moriyama\inst{1,2,3}\thanks{e-mail: moriyama@iaa.es}
  \and Alejandro Cruz-Osorio\inst{4}\thanks{email:aosorio@astro.unam.mx}
  \and Yosuke Mizuno\inst{5,6,2}
  \and Indu K. Dihingia\inst{5}
  \and Akhil Uniyal\inst{5}
}
\institute{
  Instituto de Astrofísica de Andalucía-CSIC, Glorieta de la Astronomía s/n, E-18008 Granada, Spain
  \and Institut f\"ur Theoretische Physik, Goethe-Universit\"at Frankfurt, Max-von-Laue-Strasse 1, D-60438 Frankfurt am Main, Germany
  \and Mizusawa VLBI Observatory, National Astronomical Observatory of Japan, 2-12 Hoshigaoka, Mizusawa, Oshu, Iwate 023-0861, Japan
  \and Instituto de Astronom\'{\i}a, Universidad Nacional Aut\'onoma de M\'exico, AP 70-264, Ciudad de M\'exico 04510, M\'exico
  \and Tsung-Dao Lee Institute, Shanghai Jiao Tong University, No.1 Lisuo Road, Shanghai, 201210, People's Republic of China
  \and School of Physics and Astronomy, Shanghai Jiao Tong University, 800 Dongchuan Road, Shanghai, 200240, People's Republic of China
}

\titlerunning{}
\authorrunning{Moriyama et al.}

\abstract{
The Event Horizon Telescope (EHT) has revealed the horizon-scale radiation of Sagittarius A* (Sgr A*), our galaxy's central supermassive black hole, offering a new platform to test gravitational theories. The next step involves studying accretion flows and spacetime structures near black holes using EHT time variability data and GRMHD simulations.
We study accretion dynamics in spherically symmetric black hole spacetimes deviating from general relativity, using 2D GRMHD simulations with Rezzolla-Zhidenko spacetime. 
This study systematically investigates how light curve variability amplitudes from non-Kerr GRMHD simulations depend on Schwarzschild spacetime deviations, based on the constraints from weak gravitational fields and Sgr A*'s shadow size.
We find that the dynamics of accretion flows systematically depend on the deviation.
In spacetimes with a deeper gravitational potential, fluid and Alfv\'en velocities consistently decrease relative to the Schwarzschild metric, indicating weaker dynamical behavior.
We also examine the influence of spacetime deviations on radiation properties by computing luminosity fluctuations at 230 GHz using general relativistic radiative transfer simulations, in line with EHT observations.
The amplitude of these fluctuations exhibits a systematic dependence on the deviation parameters, decreasing for deeper gravitational potentials compared to the Schwarzschild metric.
These features are validated using one of the theoretically predicted metrics, the Hayward metric, a model that describes nonsingular black holes.
This characteristic is expected to have similar effects in future comprehensive simulations that include more realistic accretion disk models and electron cooling in the future, potentially aiding in distinguishing black hole solutions that explain the variability of Sgr A*.
}



\maketitle
\section{Introduction}\label{sec:intro}
Investigating the horizon-scale emission near the black hole is crucial, offering essential observational insights into the black hole spacetime and serving as a fundamental test of general relativity.
Recently, the Event Horizon Telescope (EHT) Collaboration released the first horizon-scale images of the supermassive black holes at our galaxy's center, Sagittarius A$^*$ ({\sgra}), and in the nearby radio galaxy, M87 (\citealt{EHT_M87_PaperI, EHT_SgrA_PaperI}).
Each image shows ring morphology with diameters and structures that are consistent with the predictions of general relativity.
Notably, {\sgra}, the main target in this manuscript, is one of the most promising sources among all EHT targets due to its unique properties: 1) it has the largest apparent size of the event horizon ($\sim 50\, \uas$) (\citealt{EHT_SgrA_PaperIII, EHT_SgrA_PaperIV}), 2) its comparatively low mass produces rapid time variability events (e.g., IR/X-ray flares) with orbital timescales as short as 4-30 minutes, depending on the black hole spin (e.g., \citealt{Baganoff2001, Marrone2008, Dodds-Eden2009, Gillessen:2009, Yusef-Zadeh2009, Neilsen2013}), and 3) the precise a priori measurements of its mass and distance (\citealt{Do2019, Gravity2019}).

The radius and asymmetry of the black-hole shadow provide insights into the spacetime structure in the vicinity of the black hole.
According to general relativity, the structure of the black hole spacetime is described by the Kerr metric, characterized by the mass $M$ and the normalized black hole spin (\citealt{Kerr1963}). 
The size and asymmetry of the black hole shadow are determined by the spacetime geometry, the black hole's distance from Earth, and the viewing angle, with various alternative gravity theories offering precise solutions for the shadow's morphology (e.g., see \citealt{Younsi2016, Vagnozzi2022}).
In particular, \citet{EHT_SgrA_PaperVI} has provided state-of-the-art results for observationally constraining deviations from the Schwarzschild metric (Kerr metric for a non-spinning black hole) using {\sgra} horizon-scale imaging, thereby establishing the confidence regime for the deviation parameters.
Furthermore, \citet{Vagnozzi2022} provides a comprehensive survey considering a wide range of well-motivated deviations from black hole solutions of general relativity.
These theoretical milestones have led to a continuous series of related papers exploring the research on black hole shadows and alternative spacetime (e.g., \citealt{Cruz2021, Walia_2022, Pantig_2023, Sengo_2023, Uniyal_2023, Kocherlakota_2024, Ovgun_2024, Paul_2024}).

Despite recent observational and theoretical progress, accurately measuring the {\sgra} black hole spacetime requires achieving additional main milestones: improving the observational accuracy of the ring radius, and constructing theoretical simulations whose radiation variability is consistent with that of the {\sgra}.
According to general relativity, the diameter of the black hole shadow is predicted to depend weakly on the observer's inclination angle and the black hole spin, approximately ${\cal D}=5.0\pm 0.2\,\rg$, where $\rg = GM/c^2$ is the gravitational radius, $G$ is the gravitational constant, and $c$ is the speed of light (e.g., see \citealt{Takahashi2004, Chan2013,EHT_SgrA_PaperVI}). 
While the radius measurement has provided constraints on the parameters of each black hole spacetime according to some alternative theories, the slight difference in the shadow's size compared to predictions from general relativity remains challenging to detect given the current angular resolution of EHT observations (approximately 20\,{\uas}, equivalent to around $4\rg$).
Currently, continuous research is exploring possible methodologies to achieve accurate shadow size estimates and to place tight constraints on spacetime, especially focusing on expected future observational improvements, such as planned EHT observations, next-generation EHT (ngEHT), and the Black Hole Explorer (BHEX) projects (e.g., \citealt{Gralla_2019, Himwich_2020, Johnson_2020, Gurvits2022, Palumbo_2022, Chael_2023, Kocherlakota_2024}).

\citet{EHT_SgrA_PaperV} compared the {\sgra} data observed by the EHT in 2017, along with data from other wavelengths (86\,GHz, 136\,THz, X-rays), with theoretical models based on General Relativistic Magnetohydrodynamic (GRMHD) simulations and General Relativistic Radiation Transfer (GRRT) simulations.
They conducted theoretical GRMHD simulations, including the Magnetically Arrested Disk (MAD; e.g., \citealt{Igumenshchev2003, Narayan2003, Tchekhovskoy2011, McKinney2012}), 
Standard And Normal Evolution (SANE; e.g., \citealt{Narayan2012,Sadowski2013c}),
and tilted disk models with various parameters (e.g., \citealt{Liska2018, Chatterjee2020}), 
as well as a stellar wind-fed models with limited parameters (e.g., \citealt{Ressler2018, Ressler2020, Ressler2020b}). 
The radiation properties were then calculated by synchrotron radiation using both thermal and non-thermal electron distributions and compared with observational data.
These sophisticated observational evidence imposes stringent constraints on the model parameters of the GRMHD simulation, though they do not meet all the constraints for {\sgra}. 
The most challenging constraint to address is the light curve variability at a frequency of 230\,GHz, which is influenced by the time variation of the accretion flow near the black hole. Most GRMHD and GRRT models provide relatively large flux variability compared with that of {\sgra}.
Therefore, it is necessary for current GRMHD models to investigate additional realistic effects and take into account previously unconsidered effects to satisfy all constraints.

Several theoretical approaches for addressing these remaining questions have been proposed, each offering unique insights into the complex dynamics of accretion flows and the underlying spacetime structures. 
One of the most important realizations in this field is the inclusion of radiative cooling, which allows for a more self-consistent treatment of radiation physics. 
This approach incorporates both particle heating through magnetic reconnection (e.g., \citealt{Takahashi2016, Chael2018, Chatterjee_2023}), which is relevant for environments such as {\sgra}, and the associated cooling processes, thereby providing a more comprehensive model of energy transfer within the system.
Among the models that have been proposed, the Stellar-wind-fed Accretion Flow Model by \citet{Murchikova_2022} stands out as it more accurately captures the observed variability in the 230\,GHz light curve compared to traditional torus-initialized simulations. 
This model better reflects the physical processes occurring near supermassive black holes, highlighting the significance of stellar wind interactions in shaping the accretion dynamics.
In addition, \citealt{ChanHo2024} introduced another promising avenue for mitigating time variability in accretion flows by examining the ion-to-electron temperature ratio in strongly magnetized conditions. 
This approach suggests that adjusting the temperature dynamics between ions and electrons can have a substantial impact on the observed variability, offering a potential mechanism for reconciling discrepancies between simulations and observations.
Another promising area of study involves exploring deviations from general relativity, as investigated by \citet{Mizuno2018} and \citet{Roder2023}. 
These studies focus on the properties of accretion flows and the simulated image morphology of Kerr and dilaton black holes, exploring how these alternative spacetimes might manifest in observational data. 
Although it is difficult to distinguish different spacetimes based solely on time-averaged image morphology, the dependence of accretion flow variations and radiative time fluctuations on deviations from general relativity has not been sufficiently explored.

In this paper, we investigate the time-dependent properties and detailed dynamics of accretion flows, and compare the results with those predicted by the Schwarzschild metric in general relativity.
We perform two-dimensional GRMHD simulations to explore how deviations from the Schwarzschild black hole spacetime affect the temporal variations of accretion flows and radiation. 
Specifically, we focus on a modified spacetime where the radial and time components of the metric differ from those of the Schwarzschild solution, creating a simulation environment to study the dynamics of accretion flows.
The deviation from the Schwarzschild metric is systematically introduced using the Rezzolla-Zhidenko (RZ) parameterized metric (\citealt{Rezzolla2014}), which allows for controlled variations in the spacetime geometry. 
This approach enables us to explore a range of spacetimes that could represent more complex or realistic astrophysical scenarios, such as those involving alternative theories of gravity or modifications near the event horizon. 
Subsequently, we incorporate GRRT (General Relativistic Radiative Transfer) simulations, which are set up assuming an azimuthally symmetric distribution of the accretion flow derived from the two-dimensional GRMHD models. 
This combined approach allows us to investigate how radiation variability, a critical observational feature, responds to the underlying spacetime structure, providing insights into the potential observational signatures of these deviations across different spacetimes.

The plan of this paper is as follows. 
In Section~\ref{sec:method}, we introduce the spacetime considered in this paper and provide the setup for the two-dimensional GRMHD and GRRT simulations.
In Section~\ref{sec:grmhd_results}, we show the accretion flow property depending on the deviation from the Schwarzschild metric. 
Then, we investigate the shadow morphology and emission variability across different types of spacetimes and provide the relationship between the deviation and accretion flow variables in Section~\ref{sec:grrt_results}. 
Finally, Section~\ref{sec:summary} dedicated to summarizing the results and future prospects with the numerical simulations and expected future EHT observations.

\section{Methods}\label{sec:method}
Our investigation of the dynamics of the accretion flow around a black hole employed two-dimensional GRMHD simulations using the numerical code {\bhac} (\citealt{Porth2017, Olivares2019}), which solves the GRMHD equations on Schwarzschild and RZ spacetime backgrounds expressed in spherical horizon-penetrated coordinates $(t, r, \theta)$.
Hereafter, we introduce the target spacetime, method, and initial conditions of the numerical simulations using the geometrized units, in which the gravitational constant, $G$, and the speed of light, $c$, are set to be unity and define the gravitational radius ($r_{\rm g}:=GM/c^2$, where $M$ is the black hole mass) and time ($t_{\rm g}:=r_{\rm g}/c$).

\subsection{Target spacetime}\label{subsec:metric}

\begin{figure}
\centering
\includegraphics[width=\linewidth]{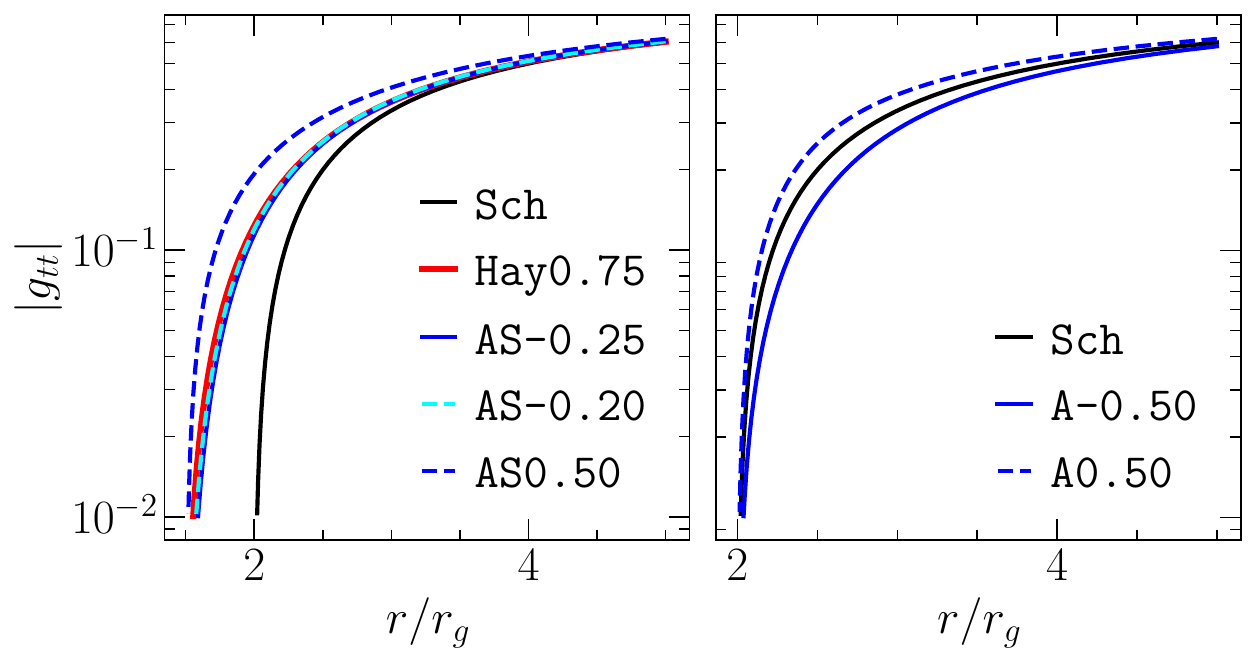}
\caption{
Comparison of the temporal spacetime metric $|g_{tt}| (=1/|g_{rr}|)$ for different horizon radii: small ($r_{0} \simeq 1.5\, r_{\rm g}$, left panel) and standard ($r_{0} = 2.0\, r_{\rm g}$, right panel). 
The black and blue curves represent the Schwarzschild metric and the RZ metric with different parameters indicated by the model symbols (see Table~\ref{tab:metric}).
The red curve illustrates the Hayward metric with a parameter $\bar{\ell} = 0.75$, and the cyan curve is the fitted metric based on the least squares method ($a_1 = -0.20, \epsilon = 1/3$).
}
\label{fig:metric}
\end{figure}

To investigate the impact of metrics that diverge from the Schwarzschild spacetime, we utilize the analytical capabilities of the Rezzolla-Zhidenko framework (\citealt{Rezzolla2014}).
This approach provides a efficient method for analyzing spherically symmetric and temporally stable black hole spacetimes across various metric theories of gravity, using a minimal set of variables as outlined in previous research.
These variables can, in principle, be obtained from near-horizon measurements of various astrophysical processes, potentially enabling more efficient tests of black hole properties and general relativity in the strong-field regime.

The RZ-parameterized spacetime is described by the line element for any spherically symmetric, stationary configuration within a spherical polar coordinate system  $(t,r,\theta,\phi)$ can be expressed as
\begin{equation*}
ds^2 = - N^2(r)dt^2+\frac{B^2(r)}{N^2(r)}dr^2 +r^2d\theta^2 +r^2\sin^2\theta d\phi^2,
\end{equation*}
where $N$ and $B$ are functions of the radial coordinate. 
The radial position of the event horizon, $r_0$, satisfies the condition $N^2(r_0)=0$.
The functions can be described with the dimensionless variable $x=1-r_0/r$ ($0\leq x\leq 1$):
\begin{eqnarray*}
N^2(x) &=& x\left[1-\epsilon (1-x) + (a_0-\epsilon)(1-x)^2 + \tilde{A}(x)(1-x)^3\right],\\
B^2(x) &=& 1+b_0 (1-x) + \tilde{B}(x)(1-x)^2,
\end{eqnarray*}
where $\epsilon:= 2\, r_{\rm g}/r_0-1$.
The functions $\tilde{A}$ and $\tilde{B}$ contribute the metric near the horizon and are finite there and at spatial infinity 
\begin{eqnarray*}
\tilde{A}(x) = \frac{a_1}{1+\frac{a_2x}{1+\frac{a_3x}{1+\cdot\cdot\cdot}}}, \ \ \ 
\tilde{B}(x) = \frac{b_1}{1+\frac{b_2x}{1+\frac{b_3x}{1+\cdot\cdot\cdot}}}, 
\end{eqnarray*}
where $\{a_i\}$ and $\{b_i\}$ are dimensionless constants.
Given the constraints derived from the Parameterized Post-Newtonian (PPN) expansion, the parameters $a_0$ and $b_0$ are determined to be negligibly small, approximately $10^{-4}$ (\citealt{Will:2006LRR}).
Consequently, this paper restricts its analysis to the scenario where $a_0= b_0= 0$.
In this manuscript, we introduce inspection metrics to analyze the impact of deviation parameters on accretion and radiation properties (Section~2.1.1).
To assess the characteristics based on the inspection metrics, we employ the Hayward metric as a representative example of a theoretically motivated spacetime (Section~2.1.2).

\subsubsection{Inspection metric}
To examine the dependence of the spacetime deviation on the different horizon radii, we introduce the inspection metrics $\texttt{A}$ and $\texttt{AS}$.
As the simplest representation, we focus on the parameters ($a_1, \epsilon$) and set $b_i=0$, {\it i. e.} ${B}^2 = 1$.
The parameters and names of the physically motivated metrics and the inspection metric are summarized in Table~\ref{tab:metric}, and the relationships of each metric are shown in Figure~\ref{fig:metric}.
For the horizon radius $r_{0}=2\rg$, a larger (smaller) $a_1$ results in a smaller (larger) magnitude of $g_{tt}$ compared to the Schwarzschild metric, corresponding to a shallower (deeper) gravitational potential (right panel of Fig~\ref{fig:metric}).
For the horizon radii $r_{0}=2.0$ and $1.5$, we introduce the ranges $-0.50 \leq a_1 \leq 0.50$ and $-0.25 \leq a_1 \leq 0.50$, respectively, within the constraints described in \citet{Cassing2023}.

\subsubsection{Hayward metric}

The static black hole solution proposed by Hayward (\citealt{Hayward2006}) replaces the singularity at $ r = 0 $ seen in the Schwarzschild black hole with a de Sitter center possessing a positive cosmological constant $ \Lambda = 3/\bar{\ell}^2 $
\begin{equation*}
g_{tt}^{\texttt{Hay}}(r) = -\left( 1 - \frac{2\,\rg \, r^2}{r^3+2\,\bar{\ell}^2 \, \rg} \right),
\end{equation*}
 where $\bar{\ell}$ is the Hubble length.
 
This model incorporates quantum corrections to address the singularity problem in standard black hole models, proposing a scenario where the central singularity is replaced with a regular finite-density core.
This metric is particularly important in the study of non-singular black holes, presenting a framework in which physical laws are maintained at the core.
This metric is expected to be justified through phenomenological approaches, such as the equation of state for high-density matter and the upper limits of curvature, as well as in quantum gravity theories. 
It could deepen our understanding of black hole physics and provide new insights into quantum gravity effects at and beyond the event horizon.
To investigate the maximum deviation of the Hayward solution from the Schwarzschild metric, we adopt an almost maximum value of $\bar{\ell} (=0.75)$, which is close to the PPN limits and corresponds to a horizon radius of $r_0 \simeq 1.5$ within the permissible range for the {\sgra} black hole solutions. This solution is described by the RZ metric with $(\epsilon, a_1, a_2, a_3, a_4) = (0.33333, -0.08333, -3.75000, 3.46667, -0.15897)$, as detailed in \citet{Kocherlakota2022}.

We summarize the used metric in Table~\ref{tab:metric}. 
Each metric has an apparent shadow radius $r_{\rm sh}$ that is approximately within the $2\sigma$ range of the {\sgra} estimates ($4.2 \lesssim r/\rg \lesssim 5.6$, \citealt{Vagnozzi2022}).
Using the least squares method, the $a_1$ value of the $\texttt{AS}$ metric, which is closest to the physically motivated $\texttt{Hay0.75}$ metric, was found to be $a_1\simeq -0.20$ (See also Fig.~\ref{fig:metric}). 
As $a_1$ of the inspection metrics increases, both the photon radius $r_{\rm ph}$ and the shadow radius $r_{\rm sh}$ monotonically increase.

\begin{table*}
\caption{Summary of model names, metric parameters, and characteristic radii.
}
\centering
\begin{tabular}{cccccc}
\hline
Metric & Symbol & Parameters & $r_0/\rg$ & $r_{\rm sh}/\rg$ & $r_{\rm ph}/\rg$ \\
\hline
Schwarzschild        & $\texttt{Sch}$ & $a_1=0.0$ & 2.0 & 5.20 & 3.00\\
Inspection metric $\texttt{A}$  &$\texttt{A}$ & $a_1=[-0.5, 0.5]$  & 2.0 & [5.58, 4.82] & [3.25, 2.81]\\
Inspection metric $\texttt{AS}$ & $\texttt{AS}$ & $a_1=[-0.25, 0.5]$  & 1.5 & [4.95, 4.42] & [2.71, 2.29]\\
\hline
Hayward              & $\texttt{Hay}$ & $\bar{\ell}=0.75$  & 1.5 & 4.93 & 2.68\\
\hline
\hline
\end{tabular}
\tablefoot{
$r_0$ is the event horizon, $r_{\rm sh}$ is the shadow size, and $r_{\rm ph}$ is the photon ring size. 
The Hubble length is denoted by $\bar{\ell}$. In the inspection models $\texttt{A}$ and $\texttt{AS}$, the parameters $b_i$ and $a_{i>1}$ are set to zero.
}
\label{tab:metric}
\end{table*}

\subsection{GRMHD simulation setup}\label{subsec:torus}

We focus our efforts on conducting two-dimensional GRMHD simulations using the RZ metric, employing the {\bhac} code (\citealt{Porth2017,Olivares2019}). 
Each simulation is conducted within a specified polar coordinate system, spanning $0.974\,r_{0}\leq r\leq 2500 \,r_g$ and $0\leq \theta \leq \pi$. 
The grid spacing in the polar coordinates is logarithmic in the radial direction and uniform in the polar direction, with a total of $(N_r, N_\theta) = (512, 256)$ grid points.
The consistency of the accretion flow is confirmed in Appendix~\ref{appendix:hiresol}.
As the initial condition for the GRMHD simulations, we introduce a hydrodynamic equilibrium torus (\citealt{Font02a}).
The inner radius of the torus is set at $r_{\rm in} = 20 \,r_g$, with the density maximum located at $r_{\rm max} = 30 \,r_g$. 
These measurements reference the horizon-penetrated RZ metric (e.g., \citealt{Mizuno2018, Roder2023}).
The specific angular momentum of the initial torus $\ell_{\rm torus}$ is approximately 5.9. 
We utilize an ideal gas equation of state with an adiabatic index of $\hat{\gamma} = 4/3$. 
To the stationary solution, we introduce a weak poloidal magnetic field, represented by a single loop and defined by the following vector potential
\begin{equation*}
A_{\phi} \propto {\rm max}(q-0.2, 0),\ \ \ q = \rho/\rho_{\rm max}.
\end{equation*}
The field strength is set such that $\beta:= 2\,p_{\rm max}/b^2_{\rm max}=100$.
In order to excite the MRI inside the torus, the thermal pressure is perturbed by 4\% white noise.
To maintain the integrity of the fluid code and circumvent issues arising from vacuum regions, we institute floor values for both the rest-mass density and the gas pressure, denoted as $\rho_{\rm fl}=10^{-5}r^{-3/2}$ and $p_{\rm fl}=10^{-7}r^{-5/2}$, respectively.
For all cells that meet the criterion $\rho\leq \rho_{\rm fl}$ we assign a value of $\rho=\rho_{\rm fl}$.
Furthermore, should any cell satisfy $p\leq p_{\rm fl}$, we set its value to $p=p_{\rm fl}$.
Based on the metrics (Section~\ref{subsec:metric}) and initial conditions of the torus (Section~\ref{subsec:torus}), we perform the two-dimensional GRMHD simulations with {\bhac} to investigate the fundamental contribution of the metric deviation from the Schwarzschild spacetime.


\subsection{GRRT simulation setup}
The radiation properties and image morphology are calculated with the general-relativistic radiative transfer code employed in the {\bhoss} (\citealt{Younsi2012, Younsi2020}), which solves the co-variant radiative-transfer equation.
In this paper, we focus on two-dimensional GRMHD simulations during the quasi-stable accretion time window, assuming a uniform azimuthal distribution to perform GRRT simulations. 
The simulated movie is calculated at a frequency of 230\,GHz, and the scaling of the mass accretion rate is set to be comparable to the mean total flux density of {\sgra} EHT observation on April 7, 2017 ($\sim 2.3$\,Jy; \citealt{EHT_SgrA_PaperI}).
The black hole mass is set at $4.1\times 10^6M_{\odot}$, and the distance is set at $8.1~{\rm kpc}$  (\citealt{Reid2009, Reid2014, Reid2019}). 
The calculation of the synthetic images is conducted using the fast-light approximation, wherein it is assumed that the fluid does not change during photon propagation, ensuring that each image corresponds to a single time slice.
The image field of view is $400\,\mu\text{as}$, and the pixel resolution is $1\,\mu\text{as}$.

Synchrotron radiation is estimated to be the primary (and sole) source of emission at 230\,GHz, with the electron energy distribution characterized by the "kappa" model (\citealt{Xiao2006}) and a non-thermal energy contribution from the jet (\citealt{Davelaar2019, Osorio2022, Fromm2022, Moriyama2024, Zhang2024}).
The electron temperature is computed through the so-called “two-temperatures” model so that the ion-to-electron temperature ratio is expressed as $T_i/T_e:=(R_{\rm low}+R_{\rm high}\beta^2)/(1+\beta^2)$ (\citealt{Moscibrodzka2009}).
We focus on the single set of the parameters $(R_{\rm low}, R_{\rm high}, \epsilon_{\kappa})=(1, 160, 0.5)$, where $\epsilon_{\kappa}$ is the fraction of magnetic energy contributing to the heating of the radiating electrons introduced in the kappa-model (e.g., \citealt{Osorio2022}). 
We introduce $\epsilon_{\kappa}$ for assessing the existence of the fainter jet component in the simulations focusing on {\sgra}.
This value offers insight into realistic jet properties, amplifying the extended jet component (e.g., \citealt{Fromm2022}).
In addition, we focus on an inclination angle of $i=30^\circ$ which is within the range of the EHT and GRAVITY constraints (\citealt{EHT_SgrA_PaperV, Gravity2020}).

\section{GRMHD simulation results}\label{sec:grmhd_results}

\begin{figure}
\centering
\includegraphics[width=\linewidth]{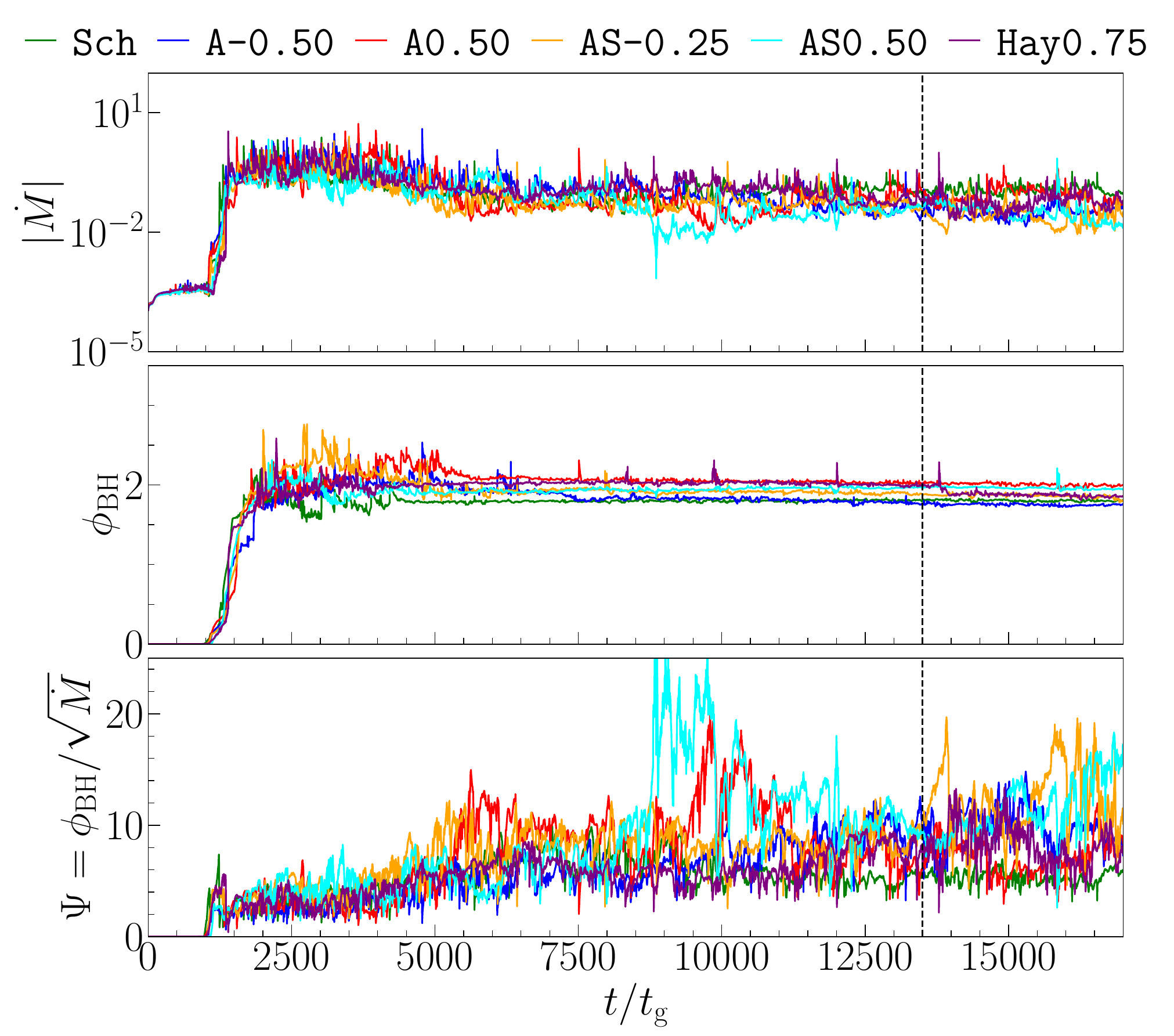}
\caption{
Mass accretion rate $\dot{M}$, magnetic flux $\phi$, and normalized magnetic flux $\Psi=\phi/\sqrt{\dot{M}}$ expressed in code units for black holes with different spacetime metrics. 
The labels of each spacetime are summarized in Table~\ref{tab:metric}. 
The time window after the vertical dotted line ($13,500 \leq t/t_{\rm g} \leq 17,000$) shows the quasi-stable period for investigating the accretion flow properties and conducting GRRT simulations.
}
\label{fig:mdot}
\end{figure}

\begin{table*}
\caption{
Time-averaged mass accretion rate, magnetic flux, and normalized magnetic flux in GRMHD simulations across different spacetimes. 
} \centering
\begin{tabular}{lccccc}
\hline
Metric & ($a_1$, $\bar{\ell}$) & $\dot{M}$ & $\dot{M}$ ($M_\odot/{\rm year}$) & $\phi_{\rm BH}$ & $\Psi$ \\
\hline
$\texttt{Sch}$ & $0.0$0 & 0.12$\pm$ 0.03 &$1.15 \times 10^{-7}-1.92 \times 10^{-7}$ & 1.80$\pm$ 0.01& 5.40$\pm$ 0.56\\
\hline
$\texttt{A}$ & -0.50 & 0.03$\pm$ 0.01 &$2.33 \times 10^{-8}-4.67 \times 10^{-8}$ & 1.75$\pm$ 0.01& 10.00$\pm$ 1.59\\
& -0.25 & 0.12$\pm$ 0.07 &$5.95 \times 10^{-8}-2.26 \times 10^{-7}$ & 1.82$\pm$ 0.01& 6.13$\pm$ 1.81\\
& 0.25 & 0.12$\pm$ 0.03 &$1.10 \times 10^{-7}-1.83 \times 10^{-7}$ & 1.88$\pm$ 0.12& 5.58$\pm$ 0.67\\
& 0.50 & 0.09$\pm$ 0.04 &$5.94 \times 10^{-8}-1.55 \times 10^{-7}$ & 2.01$\pm$ 0.19& 7.14$\pm$ 1.50\\
\hline
$\texttt{AS}$ & -0.25 & 0.03$\pm$ 0.02 &$9.80 \times 10^{-9}-4.90 \times 10^{-8}$ & 1.87$\pm$ 0.02& 12.20$\pm$ 2.94\\
& 0.00 & 0.08$\pm$ 0.03 &$6.54 \times 10^{-8}-1.44 \times 10^{-7}$ & 1.82$\pm$ 0.03& 6.95$\pm$ 1.14\\
& 0.25 & 0.06$\pm$ 0.02 &$2.49 \times 10^{-8}-4.99 \times 10^{-8}$ & 1.96$\pm$ 0.01& 8.23$\pm$ 1.24\\
& 0.50 & 0.04$\pm$ 0.03 &$8.85 \times 10^{-9}-6.20 \times 10^{-8}$ & 1.96$\pm$ 0.02& 11.40$\pm$ 2.58\\
\hline
$\texttt{Hay}$ & 0.75 & 0.06$\pm$ 0.04 &$1.29 \times 10^{-8}-6.47 \times 10^{-8}$ & 1.89$\pm$ 0.05& 8.22$\pm$ 1.74\\
\hline
\end{tabular}
\tablefoot{
The mass accretion rate $\dot{M}$ is presented in code units and $M_{\odot}/{\rm year}$, with the rescaling factor of the GRRT simulation used to match the mean total flux of approximately 2.3 Jy. The magnetic flux and normalized magnetic flux are expressed in code units (see also the main text).
}
\label{tab:accretion}
\end{table*} 

Throughout the GRMHD simulations, we continuously monitored the mass accretion rate and the magnetic flux accreting across the horizon.
Figure~\ref{fig:mdot} shows the time evolution of the mass accretion rate $\dot{M}$, the magnetic flux $\phi_{\rm BH}$, and the normalized magnetic flux $\Psi = \phi_{\rm BH}/\sqrt{\dot{M}}$ in code units for each spacetime.
\begin{eqnarray*}
\dot{M} := -2\pi \int^{\pi}_{0} \rho u^r \sqrt{-g}d\theta,\\
\phi_{\rm BH} := \pi \int^{\pi}_{0} |B^r| \sqrt{-g}d\theta,
\end{eqnarray*}
where $\rho$ is the mass density, $u^r$ is the radial component of the four-velocity of the fluid, $\sqrt{-g}$ is the square root of the negative determinant of the metric, and $B^r$ is the radial component of the magnetic field.
Both quantities are evaluated at the outer horizon $r_{0}$.
Our definition of the normalized magnetic flux $\Psi$ differs by a factor $\sqrt{4\pi}$ from the one defined in \citet{Tchekhovskoy2011}.
There is a rapid increase in $\dot{M}, \phi_{\rm BH},$ and $\Psi$ starting from $t/\tg=1,000$, indicating intensified accretion onto the black hole.
By $t/\tg=13,500$, the values of each reach a quasi-stable state.
Hereafter, we focus on the phenomena during the quasi-stable phase where $13,500\leq t/\tg \leq 17,000$ for all models.

In the quasi-equilibrium time range, the time-averaged values and standard deviations of $\dot{M}$, $\phi_{\rm BH}$, and $\Psi$ are presented in Table~\ref{tab:accretion}. 
The values of $\dot{M}$, $\phi_{\rm BH}$, and $\Psi$ in the quasi-stable time region for each spacetime are comparable. 
The values of $\Psi$ in the $\texttt{AS}$ and $\texttt{Hay0.75}$ metrics ($6.95 \leq \Psi \leq 12.00$) are slightly larger than those in the $\texttt{A}$ metrics ($5.40 \leq \Psi \leq 10.00$). 
In conclusion, the mass accretion rate and the magnetic flux accreted across the horizon do not show a clear dependence on spacetime deviations from the Schwarzschild metric, although a decrease in the horizon size slightly increases the normalized magnetic flux.


\begin{figure*}
\centering
\includegraphics[width=\linewidth]{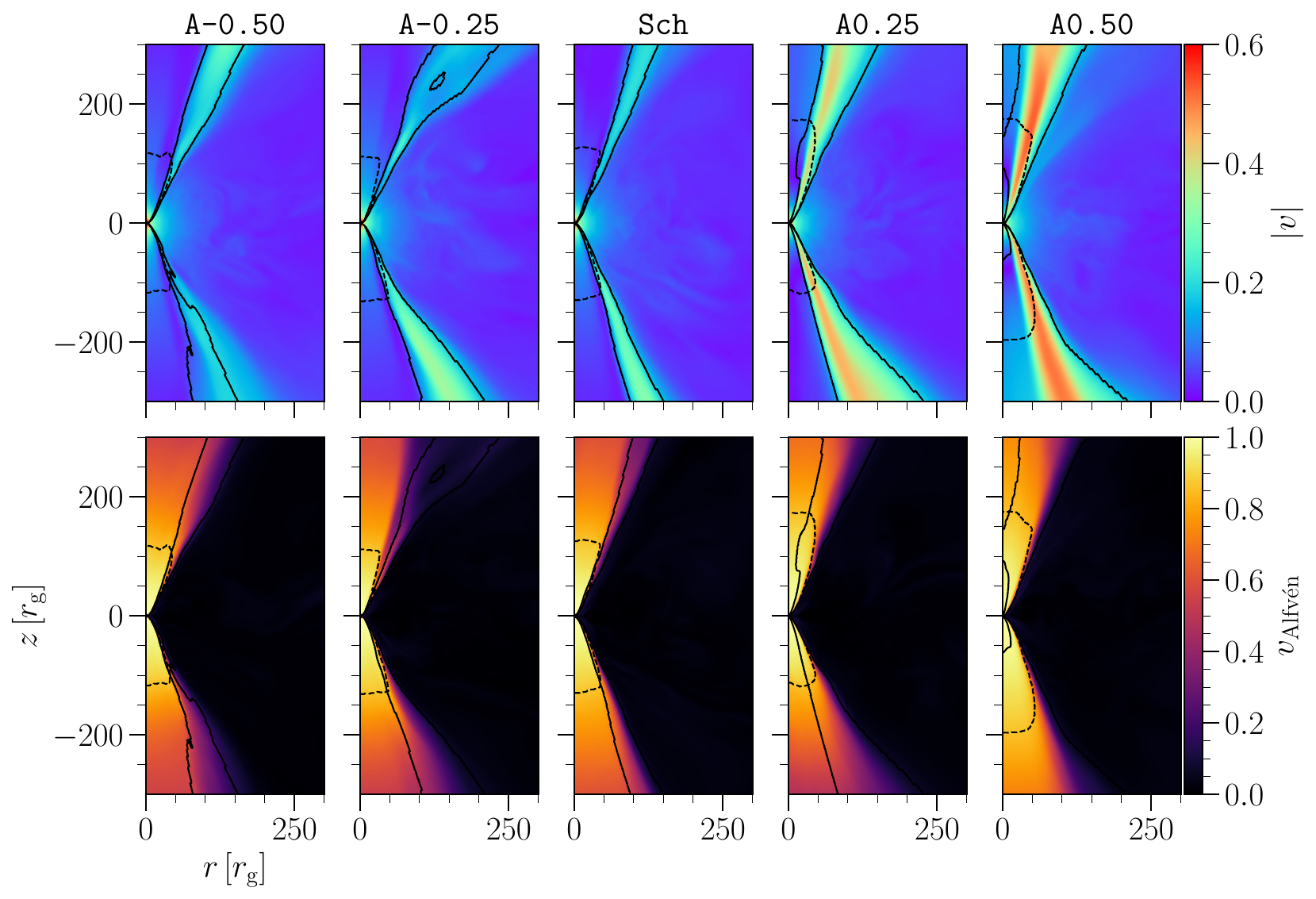}
\caption{
Time averaged distribution of the fluid and Alfv\'{e}n velocities ($|v|$ and $v_{\mathrm{Alfv\acute{e}n}}$). 
The dashed and solid black curves represent the boundary of the plasma magnetization $\sigma_{B}=3$ ($\sigma_{B}=b^2/\rho$, where $b$ is the norm of the magnetic field in the fluid frame and $\rho$ is the rest-mass density) and the Bernoulli parameter ${\rm Be} = 1.02$, respectively. 
}
\label{fig:grmhd_profile_a}
\end{figure*}

\begin{figure*}
\centering
\includegraphics[width=\linewidth]{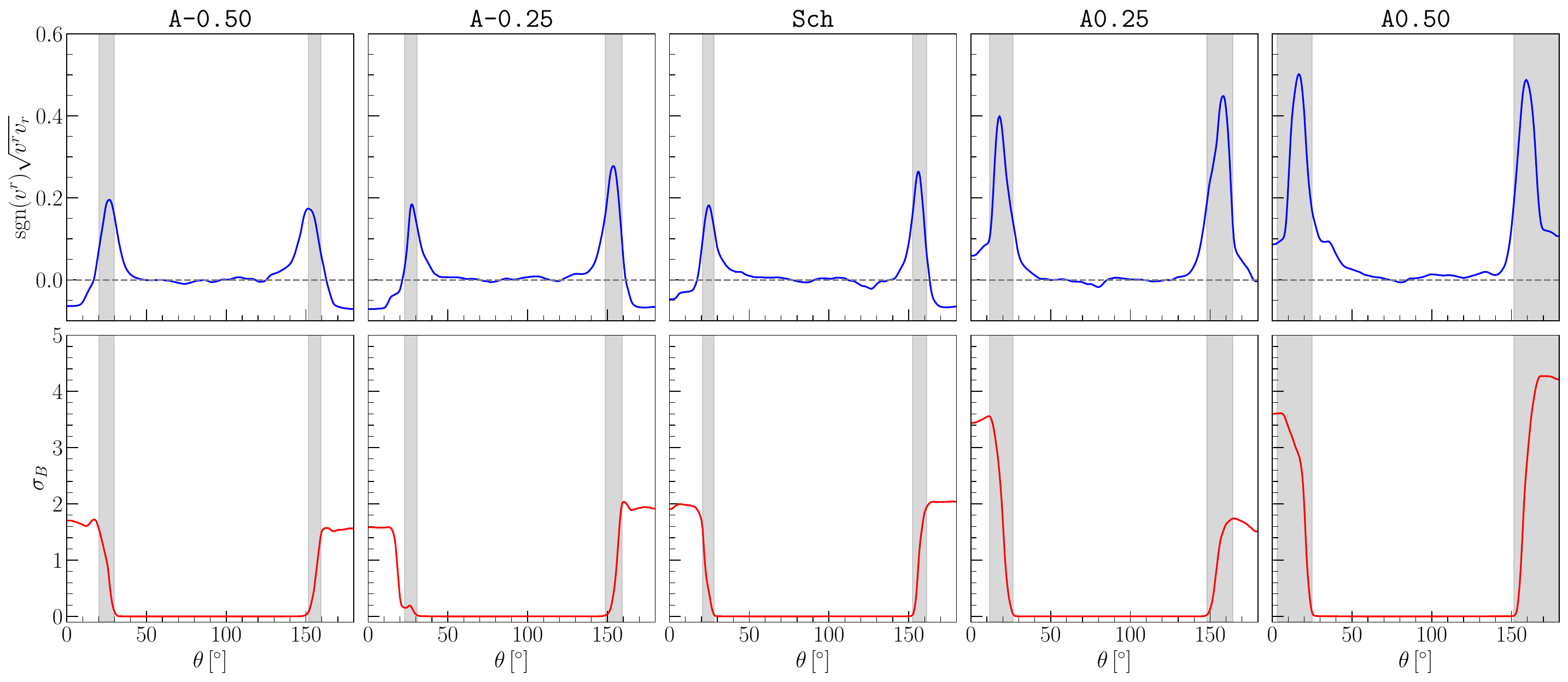}
\caption{
Polar angle distribution of the radially and time-averaged radial velocity ${\rm sgn}(v^r)\sqrt{v^r v_r}$ (top panels) and magnetization parameter $\sigma_B$ (bottom) in the range $100\leq r/\rg \leq 300$.
The gray color indicates the outflow region defined by the Bernoulli parameter ${\rm Be} > 1.02$.
}
\label{fig:vr_sigma_a}
\end{figure*}

\begin{figure*}
\centering
\includegraphics[width=\linewidth]{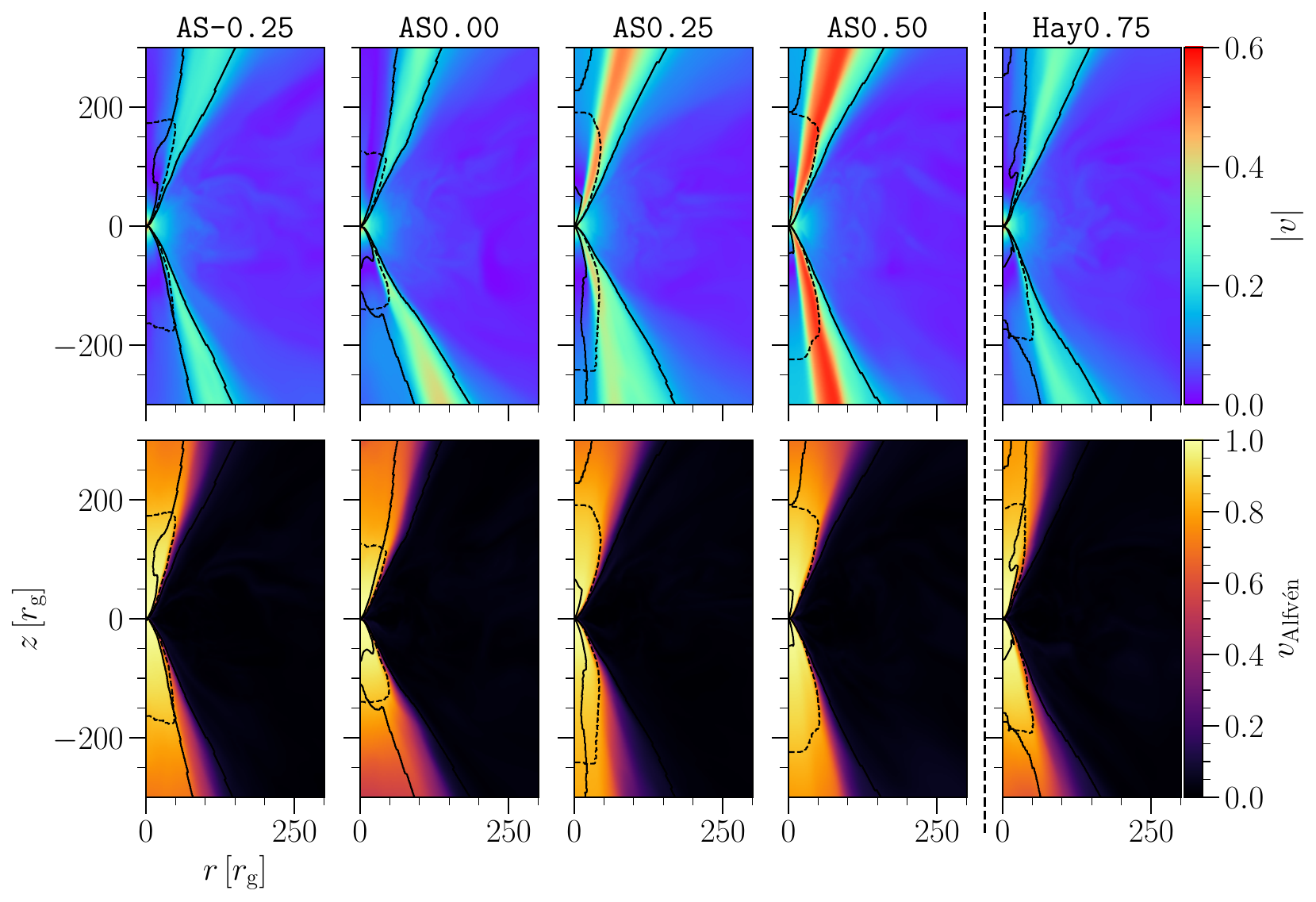}
\caption{
The same as in Fig.~\ref{fig:grmhd_profile_a}, but each profile is based on $\texttt{AS}$ and $\texttt{Hay}$ metrics. 
}
\label{fig:grmhd_profile_as_hayward}
\end{figure*}

The velocity field of the accretion flow near the black hole is highly sensitive to the underlying spacetime metric. 
In the GRMHD simulations, this dependence becomes evident when examining the distribution of velocities within the accretion flow.
As shown in Figure~\ref{fig:grmhd_profile_a}, the time-averaged distributions of the fluid's three-velocity, $|v|$ ($=\sqrt{v_i v^i}$, where $v_i$ and $v^i$ are the covariant and contravariant components of the accretion flow's three-velocity; \citealt{Porth2017}), and the Alfv\'{e}n velocity, $v_{\mathrm{Alfv\acute{e}n}}$, illustrate distinct behaviors influenced by the black hole's metric.
In every simulation executed, prominent outflow regions are observed both above and below the black hole, extending outward along the polar axis. 
These outflow regions are demarcated by a Bernoulli parameter ${\rm Be}>1.02$, indicating that material in these zones has sufficient energy to escape the black hole’s gravitational well.

The influence of the metric is particularly noticeable when examining the parameter $a_1$, which affects the curvature of the black hole spacetime. 
For $a_1 \geq 0$, as $a_1$ increases, stronger outflows, as more gravitational potential energy is released, resulting in higher values of both $v$ and $v_{\mathrm{Alfv\acute{e}n}}$. 
This trend is consistent across most metrics, indicating a robust relationship between the metric's curvature and the dynamics of the accretion flow. 
However, it is worth noting that for $a_1=-0.25$, the velocity field shows only minor deviations from that of a Schwarzschild black hole, suggesting that negative values of $a_1$ may have less impact on the flow’s dynamics compared to positive values.

In addition to the changes in velocity, the region characterized by $\sigma_{B}=3$\,---\,a measure of plasma magnetization\,---\, also expands as $a_1$ increases. 
This expansion indicates a growing influence of magnetic fields within the accretion flow, further influencing the dynamics near the black hole. 

The outflows in each spacetime are not influenced by frame-dragging, preventing the formation of standard Blandford-Znajek-type jets (\citealt{Blandford1977}), although a relatively weak outflow region is still present (Fig.~\ref{fig:grmhd_profile_a}).
To investigate the physical origin of the outflow, the radial velocity, and magnetization properties in each region are shown in Figure \ref{fig:vr_sigma_a}.
In this figure, the radially and time-averaged radial velocity, ${\rm sgn}(v^r)\sqrt{v^r v_r}$, and the magnetization parameter, $\sigma_B$, in the range $100 \leq r / r_g \leq 300$, are plotted as functions of $\theta$.
The behaviors of the outflow region (gray region, where ${\rm Be} > 1.02$), the disk region ($40^\circ \lesssim \theta \lesssim 150^\circ$), and the polar region ($\theta \lesssim 30^\circ$ or $\theta \gtrsim 160^\circ$) are examined.
Focusing on the top penels, the radial velocity in the disk region is nearly zero and shows no dependence on $a_1$. 
In contrast, in the outflow region, the radial velocity is positive for all $a_1$ values and increases monotonically with $a_1$. This feature is consistent with the results shown in Figure \ref{fig:grmhd_profile_a}. 
Additionally, in simulations of Schwarzschild black holes and GRMHD spacetimes with $a_1 < 0$, gas falls toward the black hole in the polar region. 
However, for $a_1 > 0$, outflows are observed. 
This is because the gravitational potential for $a_1 > 0$ is shallower compared to the Schwarzschild solution, allowing the interaction between axial flow and the outflow component to dominate over gravity.
Focusing on the bottom panels, in all models, $\sigma_B$ increases sharply at the boundary between the disk region and the outflow region, while the outflow region itself maintains relatively high magnetization parameters ($\sigma_B = 1$–$5$). 
This indicates that, in the simulations, magnetically driven outflows originate near the boundary between the disk region and the outflow region, driven by the magnetocentrifugal mechanism.

The same properties are observed in black hole metrics with small horizon sizes, as illustrated in Figure~\ref{fig:grmhd_profile_as_hayward}. 
For instance, the Hayward metric, which exhibits curvature similar to the $\texttt{AS}$ metric with $a_1=-0.20$, displays analogous features, confirming that both gravitational and magnetic effects play crucial roles in shaping the accretion dynamics in different spacetimes. 
The all outflow region has $\sigma_B=2-5$ and driven by the magnetocentrifugal mechanism.
These findings highlight the complex interplay between spacetime geometry, fluid dynamics, and magnetic fields in the vicinity of black holes.

\section{GRRT simulation results}\label{sec:grrt_results}

\begin{figure*}
\centering
\includegraphics[width=\linewidth]{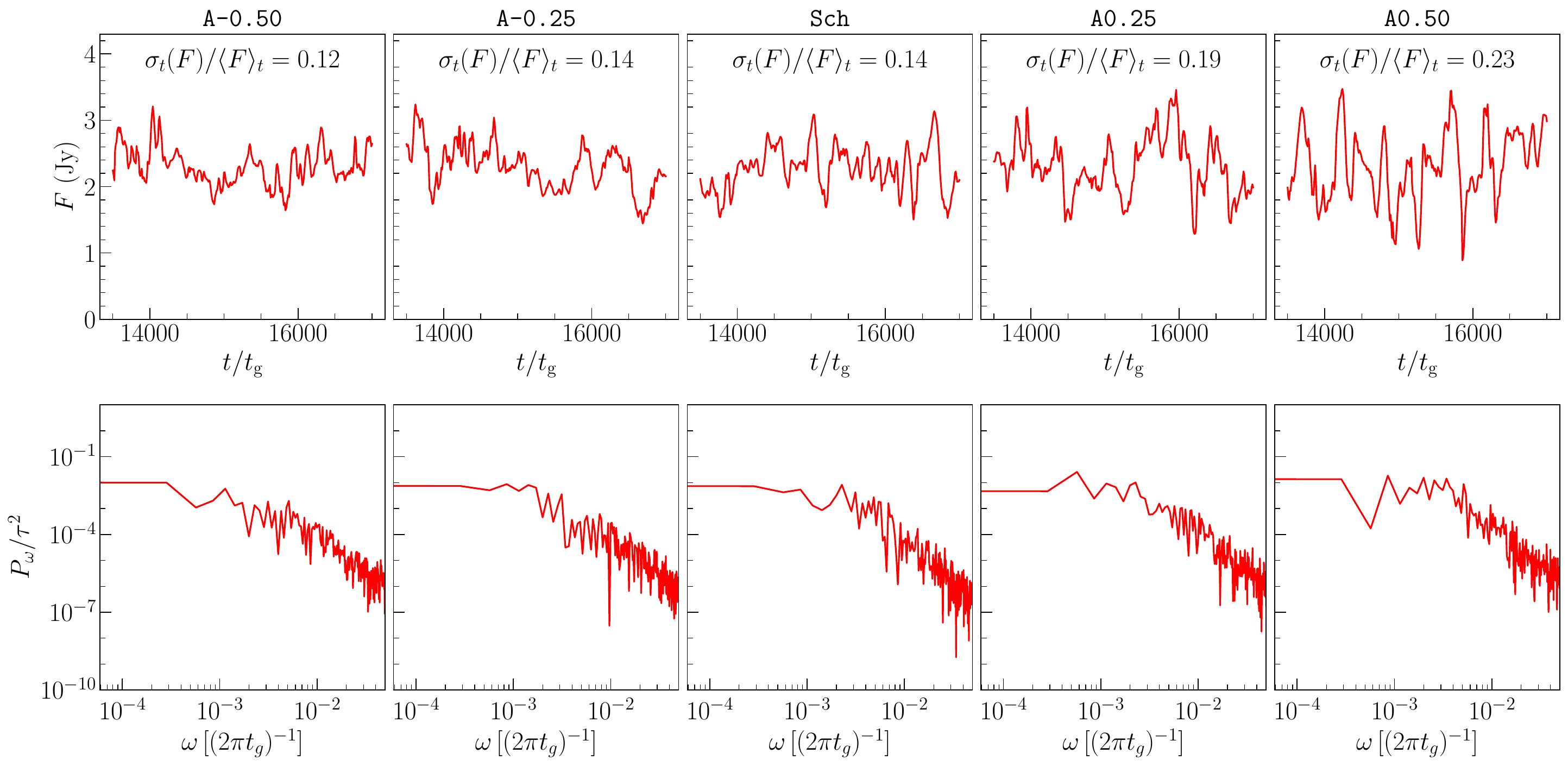}
\caption{
The spacetime dependency of the 230\,GHz light curves and power spectral densities (PSDs) is shown. 
The panels, from left to right, correspond to black holes in $\texttt{A}$ spacetimes with varying parameters $a_1$.
}
\label{fig:grrt_lc_psd_a}
\end{figure*}

\begin{figure*}
\centering
\includegraphics[width=\linewidth]{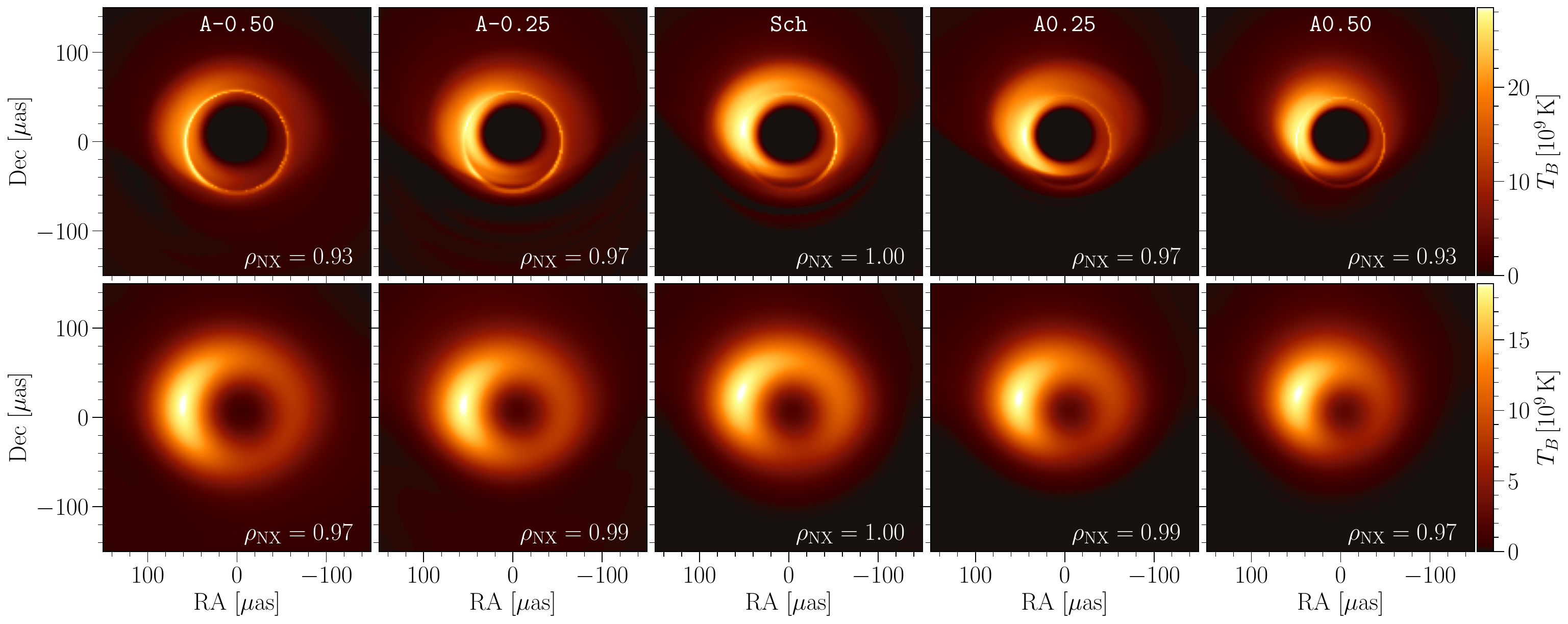}
\caption{
The metric dependence of the time-averaged 230\,GHz images (top panels) during the quasi-stable time window ($13,500 \leq t/\tg \leq 17,000$).
The bottom panels show images restored using a circular Gaussian beam with a full width at half maximum (FWHM) of 20\,\uas.
The normalized cross-correlation, $\rho_{\rm NX}$, between the Schwarzschild and each $A$ metric is shown in the bottom right corners.
}
\label{fig:grrt_image_a}
\end{figure*}

\begin{figure}
\centering
\includegraphics[width=0.8\linewidth]{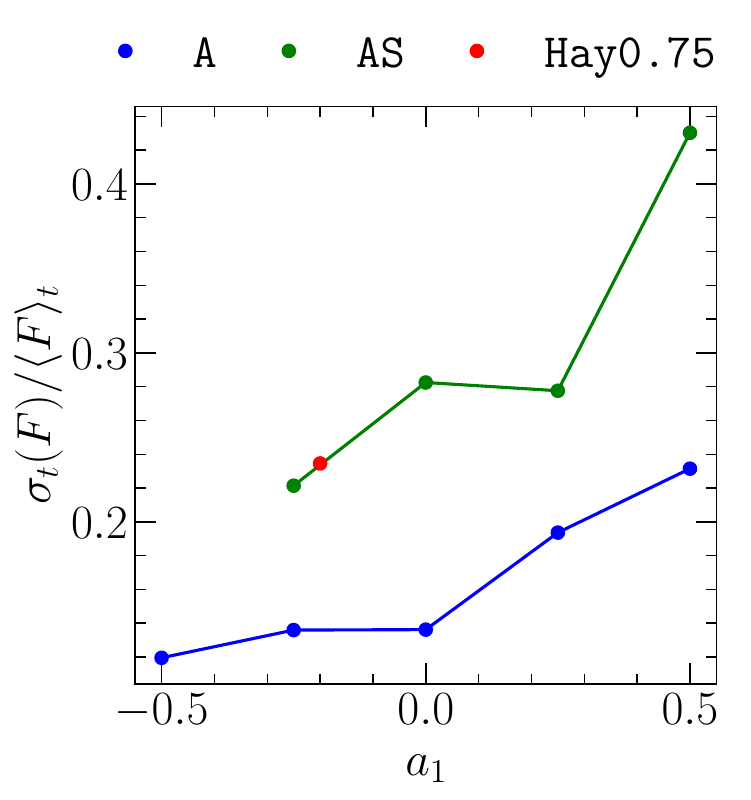}
\caption{
Summary plot of the modulation index $\sigma_t(F)/\langle F\rangle_t$.
The corresponding parameter $a_1$ of the plot of the Hayward metric is set to $-0.2$ based on the least squares method (see Section~\ref{sec:method}). 
}
\label{fig:grrt_mod}
\end{figure}

The different tendencies of the accretion flow in varying spacetimes result in systematic differences in the observed radiation. 
We report in Fig.~\ref{fig:grrt_lc_psd_a} shows the light curve and power spectrum density (PSD) for each $\texttt{A}$. 
Approximately $90\%$ of the emission comes from $r<20\,\rg$, while $50\%$ of the emission originates from $r<6\,\rg$.
The modulation index ($\sigma_t(F)/\langle F\rangle_t$, where $\langle F\rangle_t$ and $\sigma_t(F)$ is the time-averaged and standard deviation of the light curve $F$) shown at the top of each panel indicates more intense fluctuations as $a_1$ increases ($\sigma_t(F)/\langle F\rangle_t=1.2, 1.4, 1.4, 1.9$, and $2.3$ for $a_1=-0.5, -0.25, 0.0, 0.25$, and $0.5$). 
The PSDs can be characterized by a power law profile with $P_{\omega}\propto \omega^{-2.3\pm 0.5}$ at the frequency of $\omega/(2\pi t_{\rm g}^{-1})=5\times 10^{-2}$. 
This property is broadly consistent with those produced by red noise ($P_{\omega}\propto \omega^{-2}$) and previous research of the GRMHD simulations with/without non-thermal particles (\citealt{Georgiev2022, Moriyama2024}).

The simulated images across different metrics exhibit a similar crescent-like structure, reflecting the spacetime structure of the emission region near the black hole. 
In the top panels of Fig.~\ref{fig:grrt_image_a}, we present the time-averaged images with different metric parameters $a_1$, while the bottom panel shows these same images convolved with the nominal EHT resolution of 20\,\uas, reflecting the expected observational limitations as a reference.
While we introduced the non-thermal contribution characterized by the kappa model, strong jet components are not detected in the simulations for each spacetime.
As the metric parameter $a_1$ increases, a subtle trend emerges: the size of the black hole's shadow gradually decreases.
The reduction in the shadow size aligns with the behavior of key radii, including the photon ring radius ($r_{\rm ph}$) and the horizon radius ($r_0$), as summarized in Table~\ref{tab:metric}.
Although the differences in image morphologies may appear visually subtle, we introduce the normalized cross-correlation, $\rho_{\rm NX}$, to provide a quantitative measure of the structural differences across various images for each $a_1$ value (e.g., \citealt{EHT_M87_PaperIV}):

\begin{equation}
\rho_{\rm NX}(I, J) = \frac{1}{N} \sum_{i} \frac{(I_i-\langle I\rangle)(J_i-\langle J\rangle)}{\sigma_I \sigma_J},
\end{equation}
where $I_i$ and $J_i$ represent the snapshot images for different models at each pixel labeled by $i$ ($1 \leq i \leq N$), and $\langle I \rangle$ and $\sigma_I$ denote the spatially averaged intensity and standard deviation, respectively.
Each normalized cross-correlation in the figure is based on the time-averaged images of each metric case, compared to that of the Schwarzschild case.
The images convolved at 20\,\uas (bottom panel) obscure minor differences other than size, with the corresponding $\rho_{\rm NX} (>0.97)$ being comparable to the $\rho_{\rm NX}$ between the ground truth Schwarzschild and the 20\,\uas convolved images $(=0.97)$. 
These results suggest that detecting these differences in the averaged images with the current EHT resolution is challenging, as indicated by previous studies (\citealt{Mizuno2018, Roder2023}).
We note slight differences in the normalized cross-correlation between the time-averaged image and each snapshot, depending on the metric. 
Specifically, $\rho_{\rm NX} = (0.97\pm 0.02, 0.97\pm 0.016, 0.97\pm 0.02, 0.96\pm 0.02, 0.96\pm 0.02, 0.95\pm 0.03)$ for $a_{1}=(-0.50, -0.25, 0.00, 0.25, 0.50)$, respectively. 
This feature also depends on variations in the azimuthal direction, indicating that future verification using 3D GRMHD simulations is necessary.

Figure~\ref{fig:grrt_mod} summarizes the spacetime dependence of the modulation index.
The modulation index increases with the growth of $a_1$ for each metric ($\texttt{A}$ and $\texttt{AS}$). 
The shadow size in the $\texttt{AS}$ metric ($4.42 \leq r_{\rm sh} \leq 4.95$) is larger than that in the $\texttt{A}$ metric ($4.82 \leq r_{\rm sh} \leq 5.58$), and the modulation index is also larger.
These results indicate that as the event horizon and photon orbit decrease in size, the modulation index tends to increase.
Furthermore, the characteristics of the Hayward metric align with this trend. 
The modulation index for $a_1$, estimated using the least squares method, and the metric $\texttt{Hay0.75}$ agree with the $\texttt{AS}$ curve.

Because the gravitational potential of the Hayward metric for $0 < \bar{\ell} \leq 0.75$ is shallower than that of the Schwarzschild metric, and the horizon radius is smaller, the modulation index of the Hayward metric is expected to range from $\sim 0.13$ to $0.24$, based on its dependency on the shadow size and $a_1$.
This range is larger than the modulation index of the 2017 {\sgra} light curves observed with the Atacama Large Millimeter/submillimeter Array (ALMA) (0.04–0.13; \citealt{Wielgus2022}), suggesting its potential to distinguish between physically motivated spacetimes.
Based on these characteristics, we find a systematic correlation between luminosity fluctuations, characterized by the modulation index, and deviations in $g_{tt}$ from the Schwarzschild spacetime, as well as differences in horizon radii.

\section{Summary}\label{sec:summary}

In this paper, we introduced the deviations from the Schwarzschild spacetime using the RZ-parameterized metric and investigate the behavior of accretion flows near the black hole through GRMHD and GRRT simulations.
The deviations in each inspection spacetime are characterized by $a_1$ of the order $r^{-3}$, and the parameters are chosen such that the shadow size falls within the range of the {\sgra} observations from the 2017 EHT data.
For $a_1 \geq 0$, as $a_1$ increases, the gravitational potential becomes shallower compared to the Schwarzschild metric, and the characteristic radii of circular rotation decrease (Table~\ref{tab:metric}).
This spacetime results in the large dynamics of accretion flow and magnetic field characterized by the higher values of $v$ and $v_{\mathrm{Alfv\acute{e}n}}$.
These dynamic characteristics of the accretion flow are reflected in the observed radiation variations, with large luminosity fluctuations obtained for larger $a_1$ values.

Another important aspect of the accretion model for {\sgra} is the behavior of relativistic jets. 
The existence of the jet in the Galactic Center has long been debated due to the lack of a clearly collimated outflow extending significant distances from the black hole.
Nevertheless, certain cavity-like features and disordered yet bipolar structures have been proposed as indirect evidence of a weak jet (\citealt{Royster2019, Yusef2020}).
Furthermore, several GRMHD models presented in \citet{EHT_SgrA_PaperV} suggest the presence of weak jet components (\citealt{Chavez2024}). 
The theory of jet power has a rich research history, often centered around the paradigm of black hole spin (e.g., \citealt{Blandford1977, Tchekhovskoy2010}) and variations or precession in the jet's orientation (e.g., \citealt{Steenbrugge2008a, Steenbrugge2008b, Steenbrugge2010}). 
In Section~\ref{sec:grmhd_results}, it is shown that the spacetime deviation from the Schwarzschild black hole can influence the strength of the magnetic field and fluid velocity in the outflow region (Fig. \ref{fig:grmhd_profile_a}). 
However, this contribution is not sufficient to produce a clear difference in jet brightness in the GRRT simulation (Fig. \ref{fig:grrt_image_a}).

The results of this research can be used to expect the magnitude of luminosity variations and the dynamics of accretion flows in black hole spacetime solutions with various physical motivations, which were not included in the simulations discussed here, by evaluating deviations from the Schwarzschild spacetime structure.
The shallower gravitational potential, like Hayward metric, compared with that of Schwarzschild will provide larger fluid and Alfv\'{e}n velocities which will provide the large radiation variation. 
Examples of black hole spacetimes with physical motivations include other regular black holes like Hayward's, black holes in modified gravity theories, black holes with additional matter fields, black hole mimickers, exotic compact objects, and black holes with modifications from fundamental physics (see \citealt{Olivares2020, Cruz-Osorio2021, Vagnozzi2022, Chatterjee2023, Jiang2024}).

The systematic dependence of accretion flow dynamics and flux variations on deviations from general relativity is expected to persist as more realistic simulation environments are introduced.
To further advance this research and enable detailed comparisons between observational data and theoretical simulations, incorporating the azimuthal characteristics of accretion flows through 3D GRMHD simulations and carefully accounting for non-thermal influences is essential (e.g., \citealt{Meringolo2023, Imbrogno2024}).
Investigating luminosity variations within consistent torus models for Kerr black holes could provide valuable insights into whether these observational features are less dependent on specific accretion disk models (e.g., \citealt{Cruz2020, Lahiri2020, Gimeno-Soler2024, Uniyal2024}).
In addition to the above investigations, we plan to explore the effects of deviations in the $g_{t\phi}$ component, which characterizes the frame-dragging effect of a rotating black hole (e.g., \citealt{Kocherlakota2023, Ma2024}).
The continued development of these simulations, combined with the comparison of statistical observational data on the variability of Sgr A*, has the possibility to contribute significantly to distinguishing between different black hole solutions.

\begin{acknowledgements}
We thank Prashant Kocherlakota for their insightful comments on this research.
This research is supported by the European Research Council for the Advanced Grant “JETSET: Launching, propagation and emission of relativistic jets from binary mergers and across mass scales” (Grant No. 884631).
ACO gratefully acknowledges ``Ciencia Básica y de Frontera 2023-2024" program of the ``Consejo Nacional de Humanidades, Ciencias y Tecnología" (CONAHCYT, México), projects CBF2023-2024-1102 and SNII 257435. YM is supported by the National Key R\&D Program of China (grant no. 2023YFE0101200), the National Natural Science Foundation of China (grant no. 12273022), and the Shanghai Municipality Orientation Program of Basic Research for International Scientists (grant no. 22JC1410600).
The work at the IAA-CSIC is supported in part by the Spanish Ministerio de Econom\'{\i}a y Competitividad (grants AYA2016-80889-P, PID2019-108995GB-C21, PID2022-140888NB-C21), the Consejer\'{\i}a de Econom\'{\i}a, Conocimiento, Empresas y Universidad of the Junta de Andaluc\'{\i}a (grant P18-FR-1769), the Consejo Superior de Investigaciones Cient\'{\i}ficas (grant 2019AEP112), and the State Agency for Research of the Spanish MCIU through the ``Center of Excellence Severo Ochoa" grant CEX2021-001131-S funded by MCIN/AEI/ 10.13039/501100011033 awarded to the Instituto de Astrof\'{\i}sica de Andaluc\'{\i}a.
\end{acknowledgements}

\bibliographystyle{aa}
\bibliography{aeireferences}

\begin{appendix}
\section{Confirmation test with high resolution GRMHD simulation}\label{appendix:hiresol}

\begin{figure}
\centering
\includegraphics[width=0.8\linewidth]{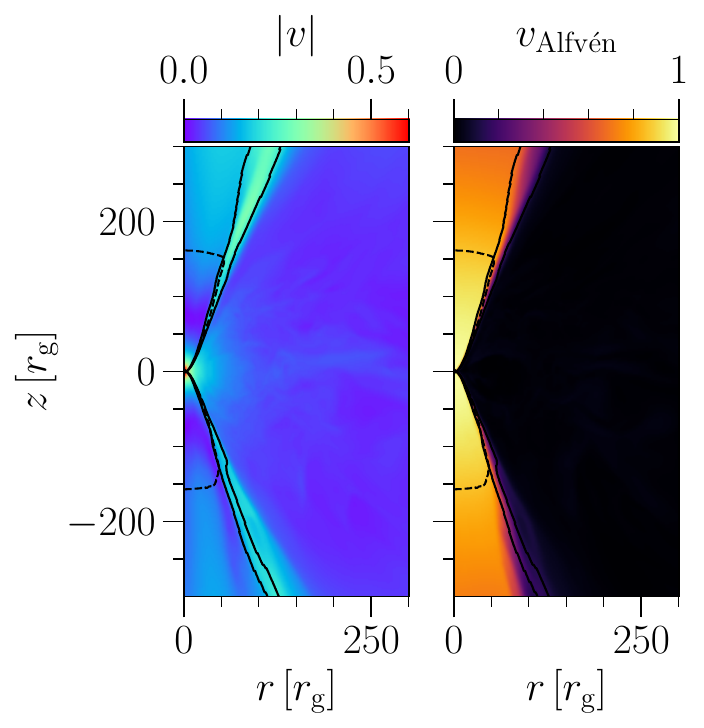}
\caption{
The same as in Fig.~\ref{fig:grmhd_profile_a}, but based on the higher spatial resolution $(N_r, N_\theta)=(1024, 512)$ compared with that in the main text $(N_r, N_\theta)=(512, 256)$.}
\label{fig:grmhd_profile_hiresol}
\end{figure}

In this paper, we discussed in detail the behavior of accretion dynamics and the characteristics of luminosity variations in the spacetime near a black hole, based on specific simulation resolutions.
To investigate the extent to which the dynamics of accretion flow depend on simulation resolution, we conducted simulations at multiple resolutions using the Schwarzschild metric.
By systematically evaluating the effects of resolution on the dynamics and behavior of physical quantities, we aimed to confirm the robustness and reproducibility of the simulation methods.
We perform GRMHD simulations under the same initial conditions with a high resolution of $(N_r, N_\theta)=(1024, 512)$, where the radial and azimuthal directions were doubled in resolution.
In this high-resolution simulation, it became possible to capture finer details of fluid motion and energy transport, and we examined how these results compared to those obtained with standard resolution, identifying any differences or consistencies.
Figure~\ref{fig:grmhd_profile_hiresol} shows the time-averaged spatial behavior of the fluid velocity distribution and Alfv\'{e}n velocity.
The simulation results at standard resolution discussed in the main text (Fig.~\ref{fig:grmhd_profile_a}) are consistent with those obtained in the high-resolution simulation, confirming that the dynamics of the accretion flow and the associated physical quantities exhibit similar behavior regardless of the differences in resolution.
This result indicates that physical insights into the behavior of accretion flows and the dynamics of energy transport remain fundamentally unchanged even when the simulation resolution is increased.
Thus, it is anticipated that the same approach will be effective in future high-resolution simulations and in analyzes that consider different spacetime geometries.

\section{Light curve variation with thermal electron distribution}\label{appendix:thermal}
\begin{figure}
\centering
\includegraphics[width=0.8\linewidth]{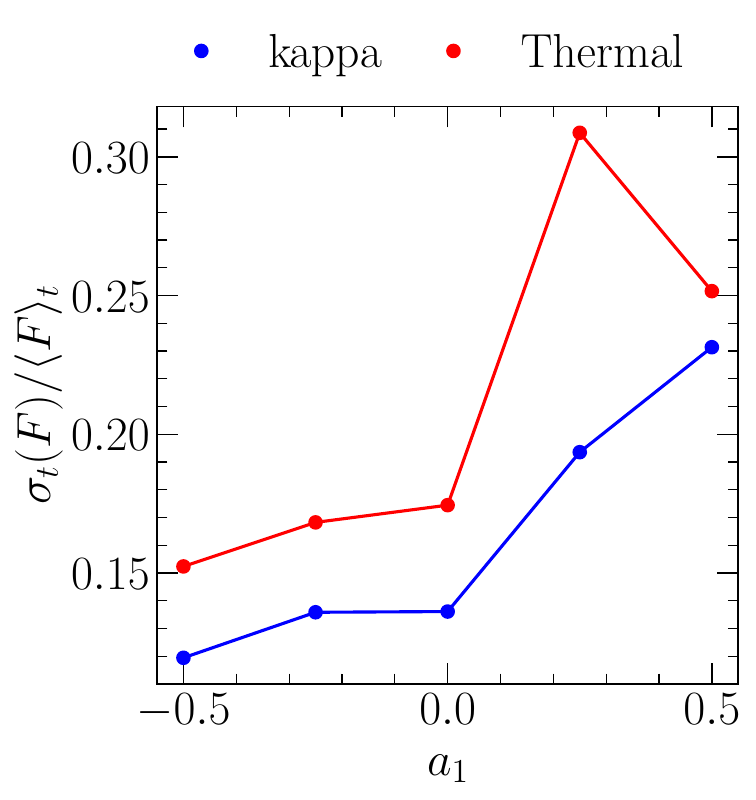}
\caption{
Modulation index $\sigma_t(F)/\langle F\rangle_t$. with the inspection metric $\texttt{A}$ based on the thermal (red) and kappa (blue) electron distribution.
}
\label{fig:grrt_mod_thermal}
\end{figure}

In the main text, we focus on the radiating electron heating mechanism introduced in the kappa model (e.g., \citealt{Osorio2022}) and investigates in detail the effects of spacetime deviation ($a_1$) on accretion flow and luminosity variability. 
This approach serves as an initial step toward understanding how the electron heating mechanism influences radiation properties and their dependence on spacetime structure in the vicinity of black holes.
Comprehensively understanding the possible range of modulation indices based on different electron heating mechanisms and initial magnetic field conditions is a key theme newly motivated by this study. 
Achieving this objective will require an extensive parameter survey, enabling deeper exploration of various physical processes in black hole environments.

In this appendix, we adopt a thermal Maxwellian distribution as an example of an alternative electron energy distribution.
The modulation indices derived from this distribution are compared with those from the kappa model, with the results detailed in Figure \ref{fig:grrt_mod_thermal}.
For this analysis, the spacetime structure of the black hole is modeled using the inspection metric $\texttt{A}$.
Additionally, the scaling of the mass accretion rate is set, as in the main text, to match the average total flux density observed for Sgr A* by the EHT on April 7, 2017 (approximately 2.3\,Jy).
For $a_1 < 0$ ($a_1 > 0$), the modulation index is smaller (larger) compared to the Schwarzschild case, a trend consistent with the results obtained using the kappa model.
Furthermore, when using the thermal model, radiation resulting from more complex accretion flow variability near the black hole tends to be more pronounced compared to the kappa model (e.g., \citealt{Fromm2022}).
As a result, the modulation indices corresponding to each $a_1$ are consistently larger for the thermal model than for the kappa model. Specifically, the modulation index for $a_1 = 0.25$ ($\sigma_t(F)/\langle F\rangle_t = 0.3$) is higher than both the value for $a_1 = 0.5$ ($0.25$) and the kappa model value ($0.23$).
The small variability observed in the 2017 {\sgra} ALMA light curve (0.04–0.13; \citealt{Wielgus2022}) suggests that the kappa model is preferable to the thermal model in spherically symmetric black hole spacetimes.

\end{appendix}

\end{document}